\RequirePackage{ifpdf}
\ifpdf 
\documentclass[pdftex]{sigma}
\else
\documentclass{sigma}
\fi

\begin{document}

\allowdisplaybreaks

\renewcommand{\PaperNumber}{014}

\FirstPageHeading

\ShortArticleName{Schr\"odinger-like Dilaton Gravity}

\ArticleName{Schr\"odinger-like Dilaton Gravity}

\Author{Yu NAKAYAMA~$^{\dag\ddag}$}

\AuthorNameForHeading{Y.~Nakayama}

\Address{$^\dag$~Berkeley Center for Theoretical Physics,
University of California, Berkeley, CA 94720, USA}
\EmailD{\href{mailto:nakayama@theory.caltech.edu}{nakayama@theory.caltech.edu}}

\Address{$^\ddag$~Institute for the Physics and Mathematics of the Universe,
University of Tokyo,\\
\hphantom{$^\ddag$}~Kashiwa, Chiba 277-8582, Japan}

\ArticleDates{Received September 16, 2010, in f\/inal form February 02, 2011;  Published online February 08, 2011}

\Abstract{We investigate possibilities for a Schr\"odinger-like gravity with the dynamical critical exponent $z=2$, where the action only contains the f\/irst-order time derivative. The Horava gravity always admits such a relevant deformation because the full ($d+1$) dimensional dif\/feomorphism of the Einstein gravity is replaced by the foliation preserving dif\/feomorphism. The dynamics is locally trivial or topological in the pure gravity case, but we can construct a dynamical f\/ield theory with a $z=2$ dispersion relation by introducing a dilaton degree of freedom. Our model provides a classical starting point for the possible quantum dilaton gravity which may be applied to a membrane quantization.}


\Classification{83D05}

\section{Introduction}
The liberation from the (local) Lorentz invariance has opened a completely new perspective of f\/ield theories and, in particular, the quantum theories of gravity. In \cite{Horava:2008ih,Horava:2009uw,Horava:2009if}, Horava proposed a~new scheme to discuss power-counting renormalizable theories of gravity based on a Lifshitz-like action with the non-relativistic dispersion relation whose dynamical critical exponent $z\neq 1$. He also proposes a relevant deformation of the theory so that the low energy ef\/fective action possesses a relativistic dispersion relation: $z=1$, and (at least superf\/icially) it recovers Einstein's general relativity at a suitable parameter point of the theory.

It is interesting to observe, however, that once the (local) Lorentz invariance is broken and if we do not impose the detailed balance condition, there would exist more relevant deformations than the Einstein--Hilbert term: it is logically possible to introduce the f\/irst order time derivative action for the metric (or any bosons).
The simplest example would be  the Schr\"odinger f\/ield theory. The Schr\"odinger action
\begin{gather}
S_S =  \int dt d^dx \left(i\Phi^* \partial_t \Phi - \frac{1}{2m} \partial_i \Phi^* \partial_i \Phi\right)    \label{ssss}
\end{gather}
has a dynamical critical exponent $z=2$ as is the Lifshitz scalar f\/ield theory with the action
\begin{gather}
S_L = \int dt d^dx \big(\partial_t \phi^* \partial_t \phi - c \partial^2 \phi^* \partial^2 \phi \big) , \label{lif}
\end{gather}
where $\partial^2 = \partial^i \partial_i$ is the spatial Laplacian.
 The f\/irst derivative term $i\phi^* \partial_t \phi$ is more relevant than the relativistic kinetic term $\partial_t \phi^* \partial_t \phi$. We try to introduce a relevant deformation $\partial_i \phi^* \partial_i \phi$ to \eqref{lif} so that we obtain the relativistic dispersion relation $z=1$ in the far infrared. However, there is no obvious reason\footnote{We could introduce the time-reversal symmetry to forbid the Schr\"odinger term. The question is whether our nature has such a symmetry.} to reject the Schr\"odinger kinetic term~$i\phi^* \partial_t \phi$ so that the infrared dynamics is not the relativistic one but the Schr\"odinger invariant system~\eqref{ssss}.

The introduction of the Schr\"odinger kinetic term drastically changes the causal structure. The non-relativistic Lifshitz action \eqref{lif} has the propagator
\begin{gather*}
\mathcal{G}_{\rm L} = \frac{1}{w^2 -cp^4} ,
\end{gather*}
and it has two poles in the energy plane that correspond to particle and anti-particle. On the other hand, the propagator for the Schr\"odinger action~\eqref{ssss} reads
\begin{gather*}
\mathcal{G}_{\rm S} = \frac{1}{w- \frac{p^2}{2m}}  ,
\end{gather*}
which has only one pole in the energy plane. Physically speaking, there is no anti-particle degree of freedom contained in the Schr\"odinger action. Mathematically, the absence of the second pole (anti-particle) improves the perturbative expansion of the Feynman diagram drastically in the Schr\"odinger f\/ield theory  because most of the vacuum diagrams now vanish due to the absence of the particle anti-particle pair creation.

In this way, once the Lorentz invariance is discarded, we may realize a new way to deform the theory so that the low energy dispersion relation has $z=2$ by introducing the f\/irst order time-derivative kinetic term. The goal of this paper is to pursue a similar possibility in theories of gravity: we would like to propose a new gravitational theory with the f\/irst order time-derivative. With the above analogy, we will call it Schr\"odinger-like gravity.
We hesitantly put ``-like" because it turns out that our theories do not have the full Schr\"odinger invariance. Unless they become topological, they do not possess the Galilean invariance and the non-relativistic special conformal invariance\footnote{We recall that a free Schr\"odinger equation has a larger symmetry than the Galilean invariance: it has a scale invariance and a non-relativistic conformal invariance in addition~\cite{Hagen:1972pd,Niederer:1972zz}.}.

As is clear from the above simple scalar f\/ield theory, we need an even number of real-valued f\/ields to construct the non-trivial Schr\"odinger-like $z=2$ dispersion relation. This will give us an obstruction to construct the gravitational analogue of the Schr\"odinger-like dispersion relation, in particular in $(1+2)$ dimension.  We will introduce a natural dilaton f\/ield to augment the additional degree of freedom needed. The introduction of the dilaton is natural in the sense that we would be able to preserve the scale invariance.

Our theory could be a non-trivial infrared f\/ixed point of the deformed Horava gravity. On the other hand, our theory by itself might give a new way to quantize gravity and for instance can be used as a new model for the dynamical gravity living on membranes. A membrane model based on the Horava gravity was proposed in~\cite{Horava:2008ih}. Similarly, we can formulate a new membrane model based on our Schr\"odinger-like dilaton gravity.

\section{Action and symmetry}\label{section2}

Our goal is to construct a gravitational action with the f\/irst order time-derivative. With the second order space-derivative in the potential term, the resulting theory would possess $z=2$ scaling around the trivial background. As we will see, this will not always be the case because the resulting theory could be locally trivial or topological. A familiar example is the Chern--Simons theory with added  magnetic f\/ield strength squared. Naive power-counting suggests a~$z=2$ scaling, but the action is rather topological because the Gauss-law constraint makes the magnetic f\/ield strength vanish. We will come back to the relation among the naive  scaling, gauge symmetry, and the triviality of the equations of motion later in Section~\ref{section2.1}.

The basic dynamical variable of the gravity is the metric. We decompose the $(d+1)$ dimensional space-time into ``time variable''~$t$ and ``spatial variables''~$x^i$ ($i=1,\dots, d$). We introduce the spatial metric $g_{ij}$ so that the line element becomes
\begin{gather*}
ds^2(t) = g_{ij}(t,x^i) dx^i dx^j  .
\end{gather*}
The line element $ds$ will a~priori depend on the time $t$. The equation of motion of our gravity system will eventually determine the time dependence of $g_{ij}$.

The line-element is invariant under the (time-independent) space dif\/feomorphism:
\begin{gather*}
\tilde{x}^i  = \tilde{x}^i(x^j),\qquad
\tilde{g}_{ij}(\tilde{x}^n)  = \frac{\partial {x}^m}{\partial \tilde{x}^i}\frac{\partial {x}^l}{\partial \tilde{x}^j} g_{ml}(x^n).
\end{gather*}
Or inf\/initesimally, $\delta x^{i} = \zeta^i(x^j)$ and $\delta{g}_{ij} = \partial_i \zeta^{k} g_{jk} + \partial_j \zeta^{k} g_{ik} + \zeta^k \partial_k g_{ij}$.
Actually, it is even invariant under the foliation preserving dif\/feomorphism (i.e.\ time-dependent dif\/feomorphism):
\begin{gather*}
\tilde{x}^i  = \tilde{x}^i(x^j,t)
\end{gather*}
by simply promoting $\zeta(x^j,t)$ to be time-dependent.
As a gravitational system, we can study the both possibilities: we will impose either the space dif\/feomorphism or the foliation preser\-ving dif\/feomorphism. The resulting Schr\"odinger-like dilaton gravity will depend on the gauge symmetry
 we impose.

\subsection{Space dif\/feomorphism vs foliation preserving dif\/feomorphism}\label{section2.1}

The dynamical content of our Schr\"odinger-like dilaton gravity crucially depends on the gauge symmetry we impose. We can either impose the space dif\/feomorphism or foliation preserving dif\/feomorphism \cite{Horava:2008ih,Horava:2009uw}. In order to realize the latter, we have to introduce additional f\/ields, lapse function $N$ and the shift vector $N_i$.
Under the foliation preserving dif\/feomorphism:
\begin{gather*}
\delta x^i = \zeta^i(x^j,t)  , \qquad \delta t = f(t)   ,
\end{gather*}
 they transform as
\begin{gather*}
\delta g_{ij}  = \partial_i \zeta^k g_{jk} + \partial_j \zeta^k g_{ik} + \zeta^k \partial_k g_{ij} + f \partial_t g_{ij},\\
\delta N_i  = \partial_i \zeta^j N_j + \zeta^j \partial_j N_i + \partial_t \zeta^j g_{ij} + \partial_t f N_i + f \partial_t N_i,\\
\delta N  = \zeta^j \partial_j N + \partial_t f N + f \partial_t N   .
\end{gather*}
We can use these new ``gauge f\/ields'' to make the action invariant under the foliation preserving dif\/feomorphism. Obviously, setting $N=1$ and $N_i=0$ will reduce the invariant action into the original one with only the space dif\/feomorphism invariance. Note that the $(d+1)$ dimensional ``metric'' may be reconstructed as \begin{gather*}
ds^2_{d+1} = -N^2 dt^2 + g_{ij}\big(N^idt + dx^i\big)\big(N^jdt+dx^j\big)   .
\end{gather*}
The $(d+1)$ dimensional metric only has a symbolic meaning because we treat the time  variable and space variables very dif\/ferently.

As is discussed in \cite{Horava:2008ih,Horava:2009uw}, there are two dif\/ferent ways to treat the lapse function $N$. One way is to treat $N$ as an arbitrary function of space and time, and integrate over the whole functional space in the path integral. In this sense, the dynamics of $N$ should be determined from the action. The other way is to f\/ix the background value of $N$ and only allow the space independent f\/luctuation in the path integral. The latter approach seems more natural because the ``gauge symmetry'' corresponding to the lapse function $N$ is time-reparametrization that does not involve any spatial coordinate\footnote{The dif\/f\/iculty to treat $N$ as an arbitrary function of space and time in the path integral is further studied in~\cite{Li:2009bg,Mukohyama:2009mz,Blas:2009yd,Henneaux:2009zb}.}. We will investigate the both possibilities in our models.

Let us see how the dynamics is generally af\/fected by the new ``gauge f\/ields'' by imposing the larger symmetry in a simpler setup. For this purpose, we consider the $z=2$ model of $(1+2)$ dimensional Abelian vector f\/ield theory. We use the obvious complex notation for the two-dimensional space: $z = x+iy$ and $\bar{z} = x - iy$ etc.
We originally have two vector f\/ields~$A$ and~$\bar{A}$ on which we impose the {\it time independent} gauge transformation: $A \to A + \partial_z \Lambda$, $\bar{A} \to \bar{A} + \bar{\partial}_{\bar{z}} {\Lambda}$, where $\partial_t \Lambda = 0$. The gauge invariant f\/ield strength is given by $F = \partial_z \bar{A} - \bar{\partial}_{\bar{z}} A$.

The gauge invariant action is given by
\begin{gather}
S = \int dt d^2x \big(iA \partial_t \bar{A} + \alpha F^2 \big)   . \label{pa}
\end{gather}
The corresponding equations of motion have a $z=2$ dispersion relation:
\begin{gather*}
-i\partial_t A + 2\alpha \partial_z(\bar{\partial}_{\bar{z}}A-\partial_z \bar{A})  = 0,\qquad
i \partial_t \bar{A} + 2 \alpha \bar{\partial}_{\bar{z}}(\partial_z \bar{A} - \bar{\partial}_{\bar{z}} A )  = 0  .
\end{gather*}

The analogue of the foliation preserving dif\/feomorphism is the time-dependent gauge transformation with~$\Lambda(t)$. The action~\eqref{pa} is no longer invariant under the time-dependent gauge transformation. To make it invariant under the enlarged gauge symmetry, we have to introduce the additional gauge connection $A_t$ that transforms as $A_t \to A_t + \partial_t \Lambda$, and  ``covariantize'' the time derivative: $\partial_t A \to \partial_t A - \partial_z A_t$. The action now reads:
\begin{gather}
S' = \int dt d^2x \big( iA\partial_t \bar{A} -iA_tF +  \alpha F^2\big)   . \label{pas}
\end{gather}

The kinetic term of~\eqref{pas} is nothing but the Chern--Simons term and the action~\eqref{pas} is invariant under the full time-dependent gauge transformation. Setting $A_t =0$ will reduce the action~\eqref{pas} to the original one~\eqref{pa}, but the equation of motion from $A_t$ gives the additional Gauss-law constraint:
\begin{gather*}
 F = 0    .
\end{gather*}
Thus, the dynamics of the foliation preserving dif\/feomorphism invariant action \eqref{pas} is (classically) trivial, and the action is topological. In a word, the additional gauge symmetry freezes the dynamics of the original $z=2$ dispersion relation.
 We will see in the following that our Schr\"odinger-like dilaton gravity reveals a similar structure.

\subsection{Pure Schr\"odinger-like gravity}\label{section2.2}

We would like to construct the Schr\"odinger-like gravity action that contains only the f\/irst order time-derivative and $z=2$ scaling symmetry. We f\/irst begin with the case when we only impose the space dif\/feomorphism. The simplest action would look like
\begin{gather}
S  = \int dt d^d x \sqrt{g} \big(g^{ij}\dot{g}_{ij} + \kappa R \big)    , \label{action1}
\end{gather}
where $i=1,\dots, d$, $\dot{g}_{ij} = \partial_t g_{ij}$ and $R$ is the Ricci scalar constructed from the spatial metric $g_{ij}$.

Although naive dimensional analysis may tell us that the equations of motion would give us a $z=2$ dispersion relation, the linearized equation does not have a $z=2$ scaling. Actually, the time derivative part of the action is rather trivial because of the identity
\begin{gather}
\partial_t{\sqrt{g}} = \frac{1}{2} \sqrt{g} g^{ij} \dot{g}_{ij}   . \label{idd}
\end{gather}
As a consequence, the f\/irst term in \eqref{action1} is a total derivative and does not contribute to the action. The other possible space dif\/feomorphism invariant scaler out of the f\/irst order time-derivative of the metric would be $\dot{g}$, but again the above identity makes it impossible to construct any non-trivial f\/irst order time-derivative action.

 The equation of motion (when $d>2$) is given by the spatial Einstein equation in $d$ dimension:
\begin{gather*}
G_{ij} = R_{ij} = 0   ,
\end{gather*}
where $G_{ij}$ is the Einstein tensor and $R_{ij}$ is the Ricci tensor: $G_{ij} = R_{ij} - \frac{1}{2}Rg_{ij}$. The absence of the time-derivative makes the time dependence of the Ricci-f\/lat metric undetermined.

Let us study a particular case of $d=2$.
In two dimension, the Einstein--Hilbert action is a~topological invariant. Therefore, one can see that the action \eqref{action1} is completely topological: the f\/irst order time derivative term is trivial and the potential term is topological. The Schr\"odinger-like pure gravity in $(1+2)$ dimension is classically a topological f\/ield theory.

Now we enlarge the symmetry of the action from the space dif\/feomorphism to the foliation preserving dif\/feomorphism. For this purpose, we introduce the lapse function $N$ and the shift vector $N_i$, and make the time-derivative in \eqref{action1} covariantized.
The covariantized action is given by
\begin{gather*}
S = \int dt d^d x \sqrt{g} \big( g^{ij}(\dot{g}_{ij} - D_i N_j -D_j N_i) + \kappa N R \big)   ,
\end{gather*}
where $D_i$ is a covariant derivative with respect to $g_{ij}$. However, the shift vector $N_i$ does not involve any dynamics because the Lagrangian involving the shift vector is a total derivative. We note that the triviality of the time-derivative has not been alleviated because the lapse function does not appear in the f\/irst order time derivative term of the action.

As in the Abelian vector f\/ield theory discussed in Section~\ref{section2.1}, typically the dynamics is more constrained with the additional gauging. The variation with respect to $N$ gives us a~``Hamiltonian constraint'':
\begin{gather*}
R = 0    .
\end{gather*}
Under the constraint, the equation of motion for the metric is
\begin{gather}
N R_{ij} + g_{ij} D^kD_k N - D_iD_j N = 0  . \label{eomm2}
\end{gather}
Taking the trace, we obtain (when $d> 1$)
\begin{gather*}
D^kD_k N = 0  , 
\end{gather*}
which demands $N R_{ij} = D_iD_j N$ from~\eqref{eomm2}.
The time dependence of the metric is arbitrary because the constraint \label{eomm} can be solved by choosing a suitable $N$ as long as we choose the trivial topology for the space.

In $(1+2)$ dimension, by using the dif\/feomorphism invariance, one can set the metric into the canonical form:
\begin{gather*}
g_{ij} = e^{\Omega(t,x^i)}g^{(0)}_{ij}    .
\end{gather*}
The f\/iducial metric $g^{(0)}_{ij}$ is f\/ixed by the complex structure, which potentially depends on $t$. We focus on the simplest topology for the two-dimensional space, so we assume the conformally f\/lat metric $g^{(0)}_{ij} = \delta_{ij}$.
 In the conformal gauge, the action reduces to
\begin{gather*}
S  = \kappa\int dt d^2x N \partial^2 \Omega   .
\end{gather*}
The equations of motion demand that $N$ and $\Omega$ are harmonic: $\partial^2 N = \partial^2 \Omega = 0$. In addition, we need to impose the Virasoro-like constraint $\partial_i \partial_j N = 0$. The time-dependence of the metric is not determined.

So far, we have treated $N$ as well as its f\/luctuation as an arbitrary function of $t$ and $x^i$. As we discussed at the beginning of this section, we can relax the condition so that $N$ only has a~spatially independent f\/luctuation. The local ``Hamiltonian constraint" is replaced by a global one:
\begin{gather*}
 \int d^d x \sqrt{g} R = 0   .
\end{gather*}
In particular, in $(1+2)$ dimension, the conformal factor is not constrained except that the Euler characteristic of the f\/iducial metric must vanish. The space as well as time dependence of $\Omega$ is completely unf\/ixed. On the other hand, the background value of the lapse function should satisfy the Laplace equation $\partial^2 N = 0$ (and stronger Virasoro-like constraint $\partial_i \partial_j N = 0$). The theory is again topological in the sense that the local equations motion do not determine the metric.

Obviously, in order to achieve the $z=2$ dispersion relation with the f\/irst order time derivative, we need, at least, two real-valued f\/ields. Otherwise, the time derivative part of the Lagrangian is a total derivative such as $h\dot{h}$.
 The failure in $(1+2)$ dimensional pure gravity for this purpose is that the dynamical degrees of freedom is just the conformal factor $e^{\Omega}$. To obtain a desired dynamical theory of gravity, therefore, we will introduce an additional scalar degree of freedom~-- dilaton in the next subsection.

\subsection{Schr\"odinger-like dilaton gravity}\label{section2.3}

The introduction of the dilaton degree of freedom in our Schr\"odinger-like gravity has two-fold purposes. The f\/irst one, as we have discussed, is to provide an extra degree of freedom to realize a $z=2$ dispersion relation. The other is to make the scaling symmetry of the action manifest. These two principles will be a guideline of the way we introduce the dilaton f\/ield $\phi$ and couple it to the metric.

We f\/irst propose the following Schr\"odinger-like dilaton gravity action:
\begin{gather}
S =\int dt d^d x \sqrt{g} e^{\phi} \big(g^{ij} \dot{g}_{ij} + \kappa R + \lambda g^{ij}\partial_i\phi \partial_j \phi \big)  .  \label{acd}
\end{gather}
This particular coupling of the dilaton makes the scaling symmetry of the action manifest
\begin{gather*}
t \to c^2 t   , \qquad x^i \to c x^i   , \qquad g_{ij} \to g_{ij}   , \qquad e^{\phi} \to c^{-d} e^{\phi}  .
\end{gather*}
The other possible term $\sqrt{g}\partial_t e^{\phi}$ is equivalent to the f\/irst term in~\eqref{acd} after integrating by part with the usage of the identity~\eqref{idd}.

Let us study the equations of motion:
\begin{gather*}
g^{ij} \dot{g}_{ij} + \kappa R -\lambda\big( g^{ij} \partial_i \phi \partial_j \phi +2g^{ij} D_i D_j \phi\big)  = 0 ,\\
-\frac{1}{2} g_{ij} \partial_te^\phi + \kappa e^{\phi} G_{ij} - \kappa \big(D_iD_j - g_{ij}D^2\big) e^{\phi} + \lambda e^{\phi}\left( D_i\phi D_j \phi - \frac{g_{ij}}{2} (D_i\phi)^2 \right)   = 0   .
\end{gather*}
It is important to observe that the Schr\"odinger-like kinetic term appears only in the trace part of the metric f\/luctuation around the trivial background $g_{ij} = \delta_{ij}(1+h)$. Thus, not all the metric components are dynamical (at least when $d>2$)\footnote{This may not be a disadvantage. From the ef\/fective f\/ield theory viewpoint in terms of the deformed Horava gravity, the non-trace part of the metric has a usual second order kinetic term. Our Schr\"odinger-like kinetic term only af\/fects the trace part (so-called the scalar mode) that should be decoupled to recover the general relativity. Our term may be used to regulate the scalar mode in the phenomenological application of Horava gravity.}. The trace mode of the metric f\/luctuation is paired with the dilaton to make the $z=2$ dispersion relation: schematically, we obtain
\begin{gather*}
\dot{h} + \kappa \partial^2 h + \lambda \partial^2 \phi  = 0,\qquad
-\dot{\phi} + \kappa \partial^2 h +\kappa \partial^2 \phi  = 0
\end{gather*}
together with the Virasoro-like constraint coming from the other component of the metric equation of motion.

In $(1+2)$ dimension, the action and the equations motion become much simpler, and the $z=2$ dispersion relation as well as the scaling symmetry of the action can be manifestly shown. For this purpose, we study the conformal mode of the metric
\begin{gather}
g_{ij} = e^{\Omega} \delta_{ij}   . \label{confc}
\end{gather}
We do not claim, unlike the case with foliation preserving dif\/feomorphism we will discuss momentarily, that  we can make the gauge choice~\eqref{confc} only with the space dif\/feomorphism. Our aim here is to study the dynamics of the conformal mode of the metric because as we have discussed, the only one scalar degree of the freedom shows a non-trivial dynamics.

The ef\/fective dynamics of the conformal mode can be encoded in the action
\begin{gather*}
S = \int dt d^2x e^\phi \big( \partial_t e^{\Omega} + \kappa \partial^2 \Omega + \lambda \partial_i\phi \partial_i\phi \big)    .
\end{gather*}
The action is invariant under the scaling transformation:
\begin{gather*}
t \to c^2 t   , \qquad x^i \to c x^i   , \qquad \Omega \to \Omega   , \qquad e^{\phi} \to c^{-2} e^{\phi}   .
\end{gather*}
The equations of motion are
\begin{gather*}
\partial_t e^{\Omega} +\kappa \partial^2 \Omega -\lambda\big( \partial_i\phi \partial_i \phi + 2 \partial^2 \phi\big)  = 0,\qquad
-e^{\Omega}\partial_t e^{\phi} +\kappa \partial^2 e^{\phi}    = 0   .
\end{gather*}
In addition, there is a Virasoro-like constraint:
\begin{gather*}
-\kappa\left(\partial_i \partial_j - \frac{1}{2}\delta_{ij} \partial^2\right) e^{\phi} + \lambda e^{\phi} \left(\partial_i\phi \partial_j \phi - \frac{g_{ij}}{2}(\partial_k\phi)^2 \right) = 0   .
\end{gather*}

By linearizing them around the trivial solution $\phi = \Omega =0$, we obtain the $z=2$ dispersion relation:
\begin{gather*}
\dot{\Omega} +\kappa \partial^2 \Omega -2\lambda\partial^2\phi  =  0,\qquad
-\dot{\phi} + \kappa \partial^2 \phi  = 0   .
\end{gather*}

If we treat $e^{\Omega}$ as a coordinate variable $q$, and $e^{\phi}$ as its canonical momentum $p$, then the ``Hamiltonian'' can be constructed as
\begin{gather*}
\mathcal{H} = -\kappa p\partial^2 \log q - \lambda p(\partial_i \log p)^2   .
\end{gather*}
The Hamiltonian always have a negative direction and the potential is unbounded, but this is a typical feature of the Euclidean--Einstein--Hilbert action.

As we have discussed, the reduced theory is close to what we would like to call ``Schr\"odinger dilaton gravity''. However, it lacks some symmetries of the full Schr\"odinger invariant f\/ield theories. In particular, it has no Galilean boost invariance, a non-relativistic special conformal invariance, and the particle number conservation~\cite{Hagen:1972pd,Niederer:1972zz}.

Now we enlarge the symmetry of the action from the space dif\/feomorphism to the foliation preserving dif\/feomorphism. As before, we introduce the lapse function~$N$ and the shift vector~$N_i$, and make the time-derivative in \eqref{action1} covariantized.
The covariantized action is given by
\begin{gather*}
S = \int dt d^d x \sqrt{g}e^{\phi} \big( g^{ij}(\dot{g}_{ij} - D_i N_j -D_j N_i) + \kappa N R + \lambda N g^{ij}\partial_i\phi\partial_j\phi \big)   ,
\end{gather*}

We f\/irst studies the constraint equations. The variation of the shift vector $N_i$ gives
\begin{gather*}
\partial_i e^{\phi} = 0   ,
\end{gather*}
which means that the dilaton $\phi$ is everywhere constant in the spatial direction. Then, the variation of the lapse function $N$ demands
\begin{gather}
 R = 0   . \label{llc}
\end{gather}

The equations of motion are derived as
\begin{gather}
g^{ij}\left(\dot{g}_{ij} -D_i N_j - D_j N_i \right) + \kappa N R  = 0 ,\nonumber\\
-\frac{1}{2} g_{ij} \partial_t e^{\phi} + \kappa N e^{\phi} G_{ij} -\kappa e^{\phi}\big(D_iD_j -g_{ij}D^2\big)N  = 0   , \label{eomful}
\end{gather}
where we have not used the constraint $R=0$ for later purposes  to recycle the same equations when the constraint \eqref{llc} is replaced by the global one. Again only the trace mode of the metric is dynamical.

As we discussed, we can relax the condition of the lapse function so that $N$ only has a spatially independent f\/luctuation. The local ``Hamiltonian constraint'' \eqref{llc} is replaced by a global one:
\begin{gather*}
 \int d^d x \sqrt{g} R = 0  .
\end{gather*}
The equations of motion \eqref{eomful} are same and they admit the $z=2$ scaling in a gauge $N_i = 0$.

Let us f\/inally specialize in the $(1+2)$ dimension. In the conformal gauge, the action is given~by
\begin{gather*}
\int dt d^2x e^{\phi}\big( \partial_t e^{\Omega} - 2\partial_i N_i + \kappa N \partial^2 \Omega + \lambda \partial_i \phi \partial_i\phi \big)  .
\end{gather*}
As in higher dimension, the variation with respect to $N_i$ and $N$ gives the constraint
\begin{gather*}
\partial^2 \Omega  = \partial_i e^{\phi} = 0   .
\end{gather*}
Thus, the dilaton has no spatial dependence at all, and the conformal factor is characterized by solutions of the Laplace equation.

The other equations of motion read
\begin{gather}
\partial_t e^{\Omega} - 2\partial_iN_i +\kappa N \partial^2 \Omega  = 0,\qquad
-e^{\Omega} \partial_t e^{\phi} +\kappa e^{\phi} \partial^2 N  = 0   . \label{eommfl}
\end{gather}
We have to also impose the Virasoro-like constraint:
\begin{gather}
\partial_i \partial_jN - \frac{\delta_{ij}}{2} \partial^2 N = 0   . \label{Virf}
\end{gather}
When $N$ is an arbitrary function, and hence $\partial^2 \Omega =0$, the f\/irst equation in~\eqref{eommfl} determines $N_i$ with respect to a given~$\Omega$, and the second equation is to be used to determine~$N$ with a given~$\phi$ under the Virasoro-like constraint~\eqref{Virf}.

On the other hand, once we relax the condition so that $N$ only has a spatially independent f\/luctuation, local ``Hamiltonian constraint'' is replaced by the global one, and in particular, it is trivial in $(1+2)$ dimension. The equations are the same as above~\eqref{eommfl} and the Virasoro-like constraint~\eqref{Virf} without $\partial^2 \Omega = 0$. It has a $z=2$ dispersion relation for small~$\Omega$ (in a gauge $N_i=0$). In this case, the spatial dependence of $N$ is f\/ixed before the equations of motion, so we regard~\eqref{Virf} as a consistency equation or tadpole cancelation condition.

As we discussed, in $d\ge 3$, the dilaton does not make all the metric mode dynamical. To avoid the problem, we can introduce the second order kinetic term \cite{Horava:2008ih,Horava:2009uw}
\begin{gather*}
\int dt d^d x \sqrt{g}N \big(K_{ij} K^{ij} -\lambda K^2\big)   ,
\end{gather*}
where $\lambda$ is a real parameter and
\begin{gather*}
K_{ij} = \frac{1}{2N} (\dot{g}_{ij} - D_i N_j -D_j N_i)   .
\end{gather*}
Apart from the f\/irst order dilaton coupling, the resultant theory is the same as the one studied by Horava. After adding the second order kinetic term, the dispersion relation of the (traceless tensor) graviton around the f\/lat background is relativistic (i.e.\ $z=1$ scaling), while the scalar mode that couples with the dilaton still satisf\/ies a f\/irst order equation. It is interesting to see if the f\/irst order dilaton coupling would solve the strongly coupled problem~\cite{Charmousis:2009tc} in the Horava gravity and can be used to modify the dispersion relation of the additional scalar mode\footnote{The author would like to thank P.~Horava for stimulating discussions. Although the mechanism works in the linearized dispersion relation, the consistency at the non-linear level seems to require a strict constraint on the solution of the equations of motion.}.

\section{Coupling with non-dilatonic matter}\label{section3}

One can couple our Schr\"odinger-like dilaton gravity with $z=2$ matter. In particular, we can study the coupling to Schr\"odinger invariant f\/ield theories.
The minimal coupling with scalars~$\Phi_I$ would be
\begin{gather*}
 S_m = \int dt d^d x \sqrt{g} \big(i\Phi_I^* \partial_t \Phi_I - g^{ij} \partial_i \Phi_I \partial_j \Phi^*_I + {\rm c.c.} \big)    .
\end{gather*}
The (de)coupling with the dilaton is f\/ixed by assigning the scaling dimension to the scalar f\/ield as
\begin{gather*}
\Phi \to c^{-d/2} \Phi   .
\end{gather*}
To make the action invariant under the foliation preserving dif\/feomorphism, we may introduce the lapse function as $ Ng^{ij} \partial_i \Phi_I \partial_j \Phi^*_I$ and covariantize the time derivative $\partial_t \Phi_I \to \partial_t \Phi_I+N^i\partial_i \Phi_I$.
We will investigate the dynamics in $(1+2)$ dimension because the situation is close to the \mbox{(non-)critical} string theory and it may have an interesting application (see~\cite{Nakayama:2004vk} for a review of the non-critical string theory and its application).

We have seen that the Schr\"odinger-like pure gravity with the space dif\/feomorphism in $(1+2)$ dimension is locally trivial. After coupling it with the Schr\"odinger matter as above, the metric equation gives the Virasoro-like constraint:
\begin{gather}
\partial_i \Phi^*_I \partial_j \Phi_I + \partial_j\Phi^*_I \partial_i \Phi_I - {g_{ij}} \partial^k \Phi^*_I \partial_k \Phi_I  + \frac{g_{ij}}{2} \left(i\Phi^*_I \partial_t \Phi_I - i\Phi_I \partial_t \Phi_I^* \right) = 0   . \label{Virrr}
\end{gather}
In the sense that the dynamics of the non-relativistic conformal f\/ield theory is constrained by the Virasoro-like condition~\eqref{Virrr}, the situation is similar to the (non-)critical string theory\footnote{It may be worthwhile noticing that the trace of the~\eqref{Virrr} does not automatically vanish unlike the string theory even if the matter sector is non-relativistically conformal.}.

We could introduce the lapse function $N$ and the shift vector $N_i$ to make the theory invariant under the foliation preserving dif\/feomorphism. The constraint coming from the shift vector $N_i$ is the vanishing of the $U(1)$ particle number current:
\begin{gather*}
i\left(\Phi_I^* \partial_i \Phi_I - \Phi_I \partial_i \Phi_I^* \right) = 0   ,
\end{gather*}
which suggests that the particle number density does not depend on $t$ (in the $N_i =0$ gauge) from the particle number conservation:
\begin{gather*}
\partial_t (\Phi^*_I\Phi_I) = 0   .
\end{gather*}

The ``Hamiltonian constraint'' from varying $N$ is more stringent. Let us work with the conformal gauge $g_{ij} = e^{\Omega} \delta_{ij}$. In the local case, we have
\begin{gather*}
\kappa \partial^2 \Omega - \partial_i\Phi^*_I \partial_i \Phi_I = 0   ,
\end{gather*}
or in the global case, we have
\begin{gather*}
 0 = \int d^2x \partial_i \Phi^*_I \partial_i \Phi_I   .
\end{gather*}
Either two cases will lead to the same conclusion
\begin{gather*}
\partial_i \Phi_I = 0
\end{gather*}
with the Euclidean signature for the target space (spanned by~$I$ indices). Therefore, no non-trivial dynamics for the matter is allowed.  It would be interesting to relax the condition by introducing a non-Euclidean signature for the target space or by allowing a non-trivial topology for the f\/iducial metric.

The condition is similar to the one we encounter in the f\/irst quantized string theory. The classical Virasoro constraint demands that the energy momentum tensor must vanish and if the target space is Euclidean, it immediately implies that there is no dynamics at all. The introduction of the time variable with the non-Euclidean signature in the target space allows us non-trivial dynamics in the string theory.

In the dilaton gravity, we f\/irst note that the simplest prescription in this section does not introduce any coupling between the dilaton and the matter. As a consequence, the equation of motion for the dilaton is unchanged.

The metric equations of motion are given by
\begin{gather*}
-\frac{1}{2}g_{ij} \partial_t e^{\phi} + \kappa N e^{\phi} G_{ij} - \kappa \big(D_iD_j -g_{ij} D^2\big)Ne^{\phi} + \lambda Ne^{\phi}\left(D_i\phi D_j\phi - \frac{g_{ij}}{2}D^m\phi D_m\phi \right) = \mathcal{T}_{ij}    ,
\end{gather*}
where the ``energy momentum tensor'' $\mathcal{T}_{ij}$ is given by
\begin{gather*}
\mathcal{T}_{ij} = -N\big(\partial_i \Phi^*_I \partial_j \Phi_I + \partial_j \Phi^*_I \partial_i \Phi_I - g_{ij} \partial_k \Phi^*_I \partial^k \Phi_I\big) - \frac{g_{ij}}{2} \big(i\Phi^*_I \partial_t \Phi_I - i\Phi_I \partial_t \Phi_I^* \big)    . 
\end{gather*}
Note that this tensor looks dif\/ferent from the conserved canonical energy-momentum tensor $T_{ij}$ of the non-relativistic scalar f\/ield theories. The latter is given by
\begin{gather*}
T_{ij} = \left( \partial_i \Phi_I^* \partial_j \Phi_I + \partial_j \Phi_I^* \partial_i \Phi_I \right) + \frac{1}{2}\left(g_{ij}\partial^2 - 2\partial_i \partial_j \right)\Phi^*_I \Phi_I   ,
\end{gather*}
where the last term is an improvement term \cite{Jackiw:1990mb}.

The constraint equation from the shift vector is given by
\begin{gather*}
2 \partial_i  e^{\phi} - i\left(\Phi_I^* \partial_i \Phi_I - \Phi_I \partial_i \Phi_I^* \right) = 0   .
\end{gather*}
Similarly, the constraint from the lapse function is given by
\begin{gather}
\kappa e^{\phi} R - g^{ij}\partial_i\Phi^*_I \partial_j \Phi_I = 0   , \label{col}
\end{gather}
for the local case, and
\begin{gather*}
\kappa \int d^2x \sqrt{g} e^{\phi} R = \int d^2x \sqrt{g}g^{ij} \partial_i\Phi^*_I \partial_j \Phi_I   ,
\end{gather*}
for the global case. In $(2+1)$ dimension, \eqref{col} suggests that $\partial_i \Phi = 0$ when $\int d^2x \sqrt{g} R = 0$, i.e. zero Euler characteristic. This can again be circumvented when the signature of the target space is not Euclidean.

\section{Discussion}

In this paper, we discussed the possibilities to construct Schr\"odinger-like dilaton gravity, where the gravitational degree of freedom has a f\/irst-order time-derivative kinetic term. The theory may be a non-trivial IR f\/ixed point of Horava gravity, where invariance under the space-time dif\/feomorphism is replaced by the foliation preserving dif\/feomorphism.

We have shown that in the pure gravity case, the inclusion of the f\/irst-order time-derivative is locally trivial, so the local dynamics of the Horava gravity cannot be modif\/ied in the IR limit by itself. However, with the additional scalar degree of freedom, we have shown that it is always possible and natural to modify the IR dynamics of the Horava gravity by including the f\/irst-order time-derivative kinetic term such as $e^\phi g^{ij} \dot{g}_{ij}$. Since the term is more relevant than the second-order kinetic term that would reproduce the Einstein action, we can never recover the Einstein gravity from the Horava gravity with an additional scalar degree of freedom once such a kinetic term is (naturally) allowed.

In reality, the candidate scalar, say, the Higgs f\/ield in the standard model is massive at our vacuum, so the allowed f\/irst-order coupling between the gravity and the Higgs f\/ield is not important in the large distance. Thus, our discussion does not exclude the Horava gravity from this viewpoint, but in the inf\/lation era, the f\/irst-order coupling between the metric and the inf\/laton f\/ield $\varphi$ would change the cosmology. For instance, in the expanding universe, $V(\varphi) g^{ij} \dot{g}_{ij}$ term would produce the ef\/fective potential for $\varphi$ as $HV(\varphi)$, where $H$ is the Hubble parameter.
We therefore suggest that whenever the inf\/lation within the Horava gravity is studied, we should take into account the f\/irst-order kinetic coupling between the metric and the inf\/laton f\/ield.

There are many possible applications. First of all, one may couple the Schr\"odinger invariant f\/ield theories to our gravitational system as we have done in Section~\ref{section3}. This will be a novel non-relativistic gravitational system coupled with the matter. One may also use our gravitational system as a basis of the membrane quantization as has been pursued in the context of Horava gravity in \cite{Horava:2008ih}. In particular, it would be interesting to study the gravitational coupling of the non-relativistic M2-brane gauge theory~\cite{Nakayama:2009cz,Lee:2009mm,Nakayama:2009ed}, which might give a new way to quantize the membrane theory in the f\/lux background.

In this paper, we have introduced the dilaton as an extra degree of freedom. As a result, the only conformal mode of the metric becomes fully dynamical. In order to make all the metric modes dynamical, we may introduce tensor degrees of freedom as a natural generalization of our approach.

Schematically, the new tensor degrees of freedom $B^{ij}$ will couple with the metric as
\begin{gather}
S = \int dt d^dx \sqrt{g} \big(B^{ij}\dot{g}_{ij} + \kappa R + K(B^{ij}) + \cdots \big)   , \label{bac}
\end{gather}
where $K(B^{ij})$ is a second-order space derivative term of $B^{ij}$ such as  $D^kB^{ij}D_{k}B_{ij}$. It is not clear what principle is needed to construct $K(B^{ij})$, but the equations of motion would look like
\begin{gather*}
\dot{B}_{ij}  = \kappa G_{ij} + \cdots,\qquad
\dot{g}_{ij}  = D^mD_mB_{ij} + \cdots  ,
\end{gather*}
where we need to specify $K(B^{ij})$ and further terms in \eqref{bac} to complete the right hand sides.

By a suitable choice of gauge (and more importantly the action itself), the linearized part of the action \eqref{bac} may be cast into the ``tensor Schr\"odinger action''
\begin{gather}
S_0 = \int dt d^d x \big( b^{ij}\partial_t h_{ij} - (\partial_k b_{ij})^2 - (\partial_k h_{ij})^2 \big) , \label{tensor}
\end{gather}
which has the full Schr\"odinger symmetry (i.e.\ Galilean invariance and the special conformal invariance as well as the particle number conservation). For this purpose, it may be useful to consider the complexif\/ied metric $\mathcal{G}_{ij} = g_{ij} + i B_{ij}$ due to the $U(1)$ rotation symmetry in \eqref{tensor}.
It is interesting to study further the natural complexif\/ied metric and the doubled geometry within the context of the Schr\"odinger gravity. We hope to come back to the issue in the near future.

\subsection*{Acknowledgements}

The work was supported in part by the National Science Foundation under Grant No.\ PHY05-55662 and the UC Berkeley Center for Theoretical Physics  and World Premier International Research Center Initiative (WPI Initiative), MEXT, Japan.

\pdfbookmark[1]{References}{ref}
\LastPageEnding

\end{document}